# Computationally-guided discovery and synthesis of the amorphous nitride $Y_2WN_4$

O.V. Pshyk[a], S. Zhuk[a], J. Patidar[a], A. Wieczorek[a], A. Sharma[b], J. Michler[b], C. Cancellieri[c], V. Stevanovic[d,*], S. Siol[a,*]

a) Empa – Swiss Federal Laboratories for Materials Science and Technology,
   Laboratory for Surface Science and Coating Technologies, Dübendorf 8600, Switzerland
b) Empa — Swiss Federal Laboratories for Materials Science and Technology,
   Laboratory for Mechanics of Materials and Nanostructures, Thun 3602, Switzerland
c) Empa – Swiss Federal Laboratories for Materials Science and Technology,
   Laboratory for Joining Technologies and Corrosion, Dübendorf 8600, Switzerland
d) Colorado School of Mines, Golden, 80401, CO, United States

*Corresponding authors: vstevano@mines.edu, sebastian.siol@empa.ch*

## Abstract

Amorphous materials offer unique functional characteristics, which are often not observed in their crystalline counterparts. This makes them invaluable for many technological applications, such as diffusion barriers in semiconductor devices. However, the computationally guided search for new functional amorphous materials with attractive properties represents a major challenge. In this work, we combine theory and experiment to discover and synthesize the amorphous ternary nitride $Y_2WN_4$. We show how computational random structure sampling offers a route to robustly identify chemistries which are hard to crystallize. Experiments prove that the predicted nitride is easily synthesized in amorphous phase with no detectable precipitates. The material exhibits remarkable stability against crystallization at high temperature and as well as excellent oxidation resistance and stability against Cu diffusion. Moreover, $Y_2WN_4$ exhibits a sharp onset of optical absorption and an indirect band gap of 2.24 eV. These properties make this material promising for the integration in electronic devices as a high-performance diffusion barrier with adjustable band edges.

# Introduction

Amorphous materials are invaluable for countless applications providing functionalities which are often not observed in their crystalline counterparts [1–3]. For instance, the unique carrier transport properties of amorphous materials allowed to develop transparent n-type [4] and high-performance p-type [5] amorphous oxide semiconductors. Moreover, amorphous materials have shown competitive electrocatalytic [2,6] and biomedical characteristics [7] owing to the abundance of unsaturated coordination sites and unique local short-range structure. Amorphous thin films often demonstrate exceptional diffusion barrier performance due to the absence of grain boundaries. This makes them most successful as diffusion barrier layers in microelectronic circuits [8,9] or corrosion and oxidation resistant coatings [10]. However, the rational design of amorphous thin films is presently largely hindered by the gaps in predictive understanding of the tendency to form amorphous phases in physical vapor deposition or other non-equilibrium synthesis techniques. Much of the research remains limited to certain material classes such as bulk metallic glasses and is not generally applicable outside of those groups [11]. In contrast to crystalline materials, very few approaches for the design of amorphous phases have been reported in the literature [12]. The typical computational methods and/or screening of computational databases for materials with desired properties is not feasible for the design of amorphous materials, since high-throughput computational efforts typically do not include disordered structures [13].

One targeted application for a new material could for instance be a thin diffusion barrier with high oxidation resistance and phase stability. For this application amorphous transition metal nitrides appear to be ideal candidates. However, group III-VI transition metals have a strong affinity to nitrogen and are known to form crystalline nitrides, even during low temperature deposition. This makes the targeted synthesis of amorphous transition metal nitride thin films challenging. Therefore, there is a need for an appropriate design approach for identifying compound systems with a strong tendency to amorphization and an attractive composition space for functional property tuning. The amorphous phase formation rules have been formulated for bulk metallic glasses [11] and high entropy alloys [14,15], all having complex multielement compositions. But the amorphous phase formation from compounds which have a preference to form strong chemical bonds, like group III-VI transition metal nitrides, remains hard to predict.

Here, we report on the discovery of new amorphous ternary nitrides within the Y-W-N chemical space and the methodology we developed that enabled it. Computationally-guided experimental screening of the YN-WN pseudo-binary system allowed us to predict and synthesize amorphous transition metal nitride thin films that are exceptionally resistant toward crystallization. The tendency of a particular chemistry within $Y_xW_{1-x}N$ system to form amorphous rather than crystalline phase is predicted using the first-principles random structure sampling method [16]. Subsequent exploratory synthesis and high-throughput characterization of $Y_xW_{1-x}N$ thin films allowed for a rapid screening of the entire composition space for amorphous phase formation.

## Computational prediction and experimental verification

We start by selecting the WN-YN pseudo-binary system for this study using simplistic amorphous phase formation rules known for ionic-covalent compounds. The most basic approach for selecting potential candidates for amorphous nitride thin films is based on the consideration of a possible miscibility between the ground state binary phases or other possible polymorphs existing for a given compound system. WN-YN is a promising system due to the different ionic cation radii and the large number of competing polymorphs in both the WN and YN system (see Supplementary note 1 and Fig S1). In addition, we hypothesize that alloying of WN and YN might lead to improved oxidation resistance not found in either of the binaries (Supplementary note 2). After selection of the material system, following the octet rule, we identify the stoichiometries potentially leading to the formation of chemically stable phases within the system are $Y_2WN_4$ and $YWN_3$ (given by $N^{3-}$, $Y^{3+}$ and $W^{6+}$). For these compositions we carry out first-principles random structure sampling to evaluate the preference to form a crystalline phase. The first-principles random structure sampling includes (a) generation of a large number of periodic random structures (random lattice parameters and atomic positions) and (b) their subsequent relaxation using density functional theory (DFT) to the closest local minimum on the potential energy surface. This provides information about the statistical/probabilistic relevance of the set of obtained local minima. It has been shown previously [16–18] that the experimentally observed metastable polymorphs in a number of systems are exclusively those that are the top occurring in the random sampling. This result can be interpreted in terms of the size ("width") of their local minima and the implied higher probability to "fall into" larger local potential wells and minima.

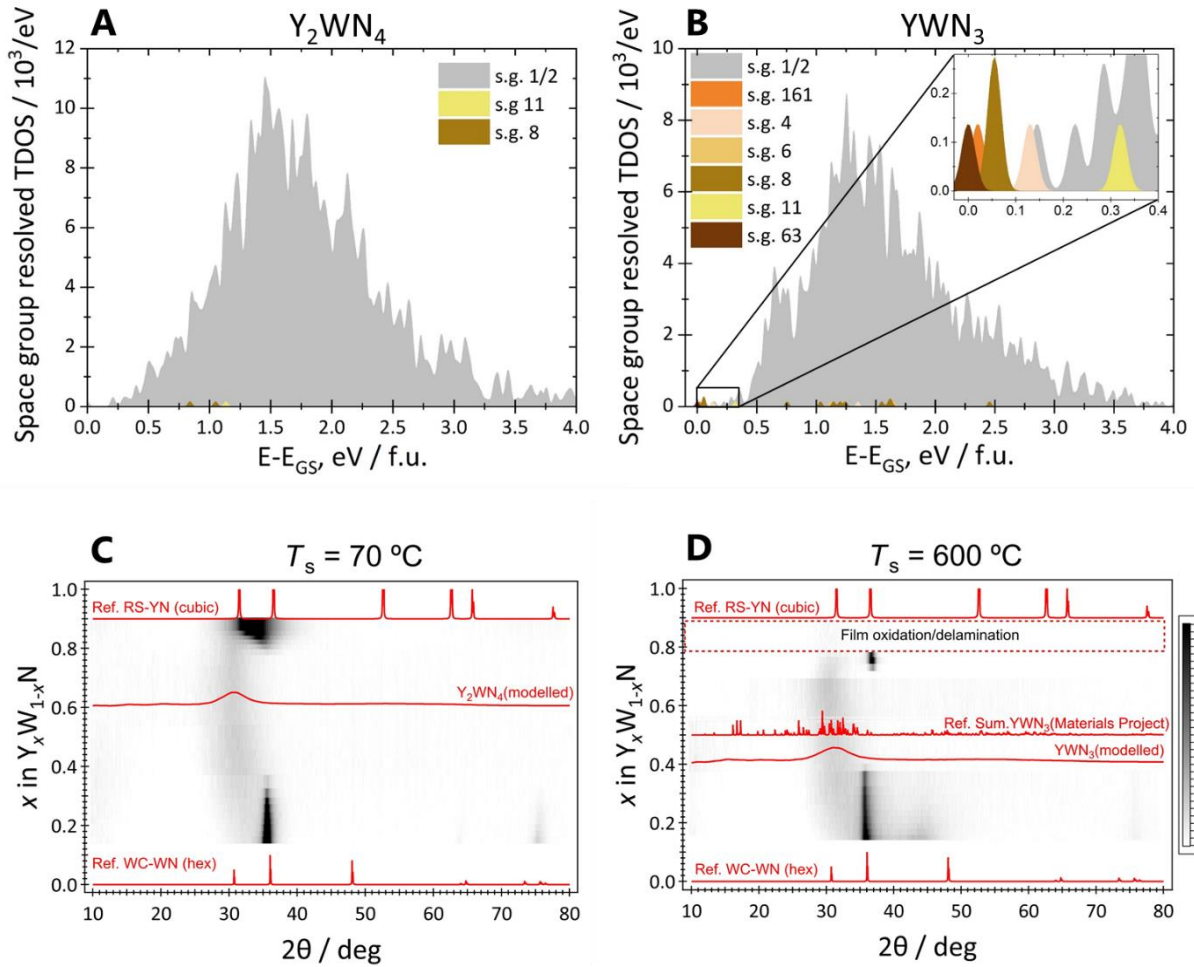

**Fig. 1 Computational structure predictions and experimental verification.** The first-principles random structure sampling: plots of the space group resolved thermodynamic density of states (TDOS) for (A) $Y_2WN_4$ and (B) $YWN_3$. Combined XRD and XRF color maps of $Y_xW_{1-x}N$ thin films synthesized at $T_s$ of (A) 70 °C and (B) 600 °C. XRD pattern of hexagonal WC-type WN and cubic NaCl-type YN are given for reference in (C) and (D). (C) shows modelled XRD pattern for $Y_2WN_4$ and (D) for $YWN_3$ as well as a simulated XRD pattern for a sum of all theoretically predicted space groups for $YWN_3$ available in the Materials Project database [13].

In this paper we validate the opposite hypothesis: Stoichiometries that do not exhibit any particularly "wide" local minimum, that is, not having any statistically preferential crystalline structure(s), are prone to easily form amorphous phases that are distinctly stable against crystallization. Results showing the energy distribution of the (DFT-relaxed) random structures are provided in Fig. 1A and 1B. The space group-resolved thermodynamic density of states (TDOS) across 2000 random structures obtained by the random sampling shows no preference for crystalline phase formation for $Y_2WN_4$ (Fig. 1A) and a marginal probability for $YWN_3$ (Fig. 1B). For $Y_2WN_4$ the only symmetric (crystalline) phases that are found belong to the space groups 8 and 11, both occurring only once. All other local minima belong to the least symmetric space groups 1 and 2, which can be viewed as disordered (non-crystalline) structures. $YWN_3$ was previously predicted to have the monoclinic ground state structure with space group 15 [19], which is not found in the random sampling, suggesting a possibility of a very narrow local minimum. We do find structures with space groups having some symmetry (4, 8, 11, 63, 161) but their occurrence is still 1 to 2 out of 2,000, which cannot be interpreted as them being more statistically significant than the majority of low symmetry structures (s.g. #1 and #2). It is important to note that these results are in stark contrast to other ternary nitrides we studied previously for which the as-synthesized cation-disordered rocksalt phase and/or various ordered ground state structures have been successfully identified in the random structure sampling [20,21]. In this context, the random sampling results for both $Y_2WN_4$ and $YWN_3$ imply an absence of statistically more significant crystalline local minima, which in turn implies high probability that these two systems will form amorphous phases upon synthesis, in particular $Y_2WN_4$.

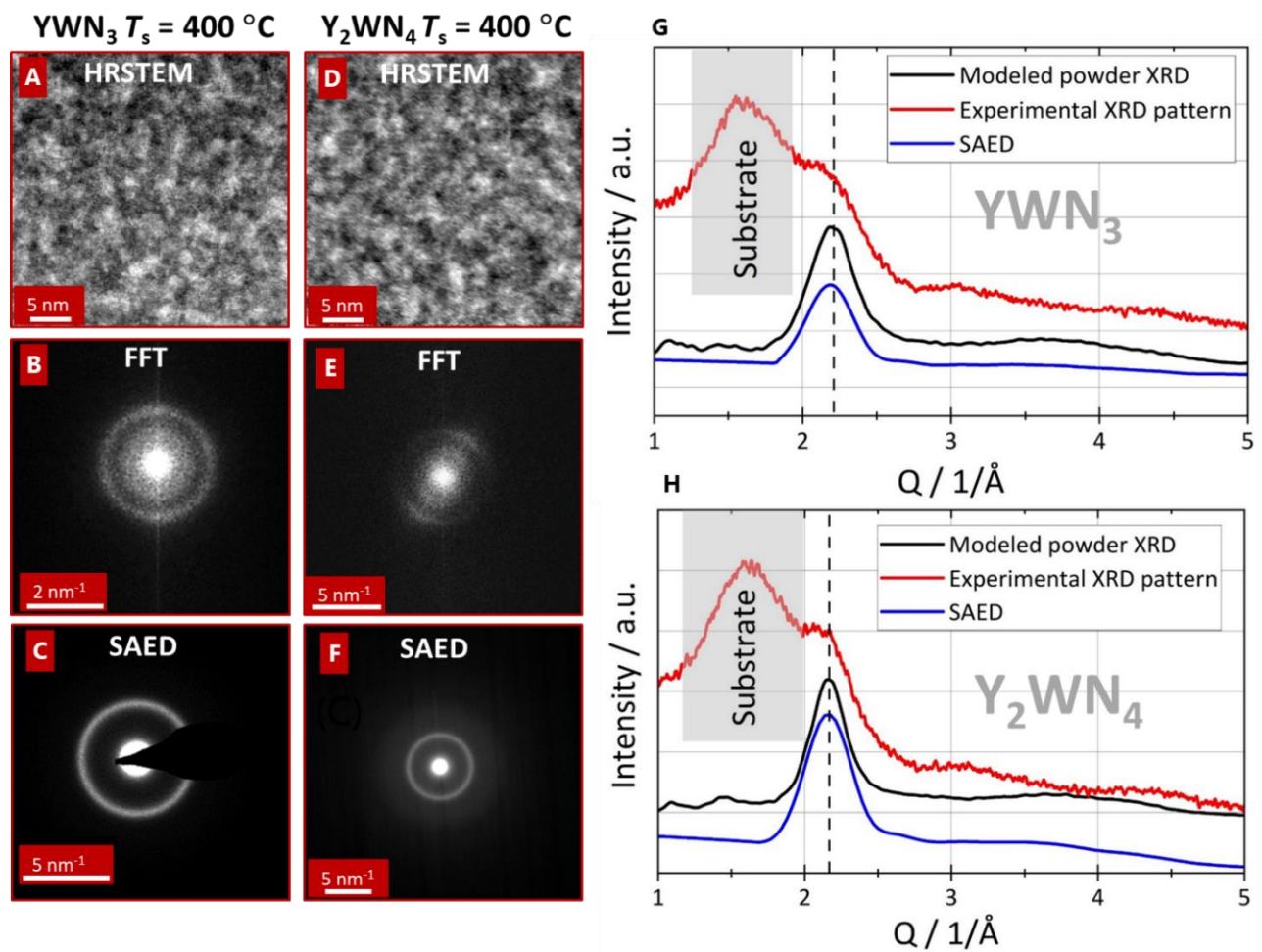

**Fig. 2 Local and global structural characterization.** HRSTEM images, corresponding FFT patterns, and SAED patterns for (A-C) $YWN_3$ films and (D-F) $YWN_3$ films grown at $T_s$ = 400 °C. All thin films are grown on borosilicate glass substrates. Comparison of theoretical and experimental reciprocal distances: Q obtained from modelled powder XRD, experimental XRD and radially integrated SAED for (G) $YWN_3$ and (H) $Y_2WN_4$ grown at $T_s$ = 400 °C. The black dashed lines depict the peak maximum of the modelled XRD peak.

To verify the latter, combinatorial physical vapor deposition (PVD) is used for a rapid screening of the $Y_xW_{1-x}N$ phase space by reactive magnetron co-sputtering of combinatorial thin film libraries covering the composition window within $0.15 \leq x \leq 0.9$, whereas the substrate temperature ($T_s$) is varied from 70 °C up to 600 °C across the substrate. High-throughput screening of the composition and structure of the libraries by means of X-ray fluorescence (XRF) and X-ray diffraction (XRD) reveals a wide composition range $0.32 \leq x \leq 0.80$ where thin films are X-ray amorphous with no signs of precipitation at $T_s = 70°C$. Moreover, the composition region of the amorphous phase stability shrinks only slightly to $0.40 \leq x \leq 0.72$ at $T_s = 600$ °C, demonstrating a remarkable temperature stability of the amorphous phase. Furthermore, the experimental XRD patterns show an excellent agreement with modeled XRD patterns (averaged over the ensemble of random structures) for both $Y_2WN_4$ and $YWN_3$ (Fig. 1C, the reciprocal interatomic distances are presented below in Fig. 2G, H) supporting the validity of the theoretical approach. The calculations also allow to model neutron and X-ray scattering structure factors for these materials (Fig. S2) showing typical patterns for amorphous materials.

Single composition samples of $Y_2WN_4$ and $YWN_3$ thin films are grown on different substrates (see Materials and Methods) for further thorough studies of these amorphous materials. The local and global structural assessment of $Y_2WN_4$ and $YWN_3$ thin films grown at $T_s = 400$ °C using high-resolution scanning transmission electron microscopy (HR-STEM) and selected area electron diffraction (SAED), respectively, reveals that all films are completely amorphous without any sign of crystallization. Fast-Fourier transform (FFT) patterns from corresponding HR-STEM images show broad halos typical for amorphous materials [22]. Moreover, the large-aperture SAED patterns acquired from both films exhibit a broad halo confirming amorphous structure and no visible precipitation of crystalline phases at a more global scale. The reciprocal interatomic distances obtained from SAED and XRD are in an excellent agreement with the values obtained from modeled XRD patterns (Fig. 2 G, H) for both $Y_2WN_4$ and $YWN_3$. The analysis of pair distribution functions (PDF) derived from SAED following an approach described elsewhere [23] demonstrates short-range order in both $YWN_3$ and $Y_2WN_4$ films grown at $T_s = 400$ °C (Fig. S3). Moreover, the lengths of W-N, Y-N, and Y-W bonds determined from SAED-derived PDFs are in good agreement with the values obtained from the theoretical total and partial geometric PDFs (Supplementary Table 1, Fig. S4). The comparison of $YWN_3$ thin films grown at different $T_s$ (Fig. S5) shows no impact of $T_s$ on the microstructure of the films. Low magnification STEM imaging, energy-dispersive X-ray spectrometry (EDS) elemental mapping and linescans show no apparent intensity fluctuations confirming chemical homogeneity at the nano- and micro-meter scale (Fig. S6, Fig. S7).

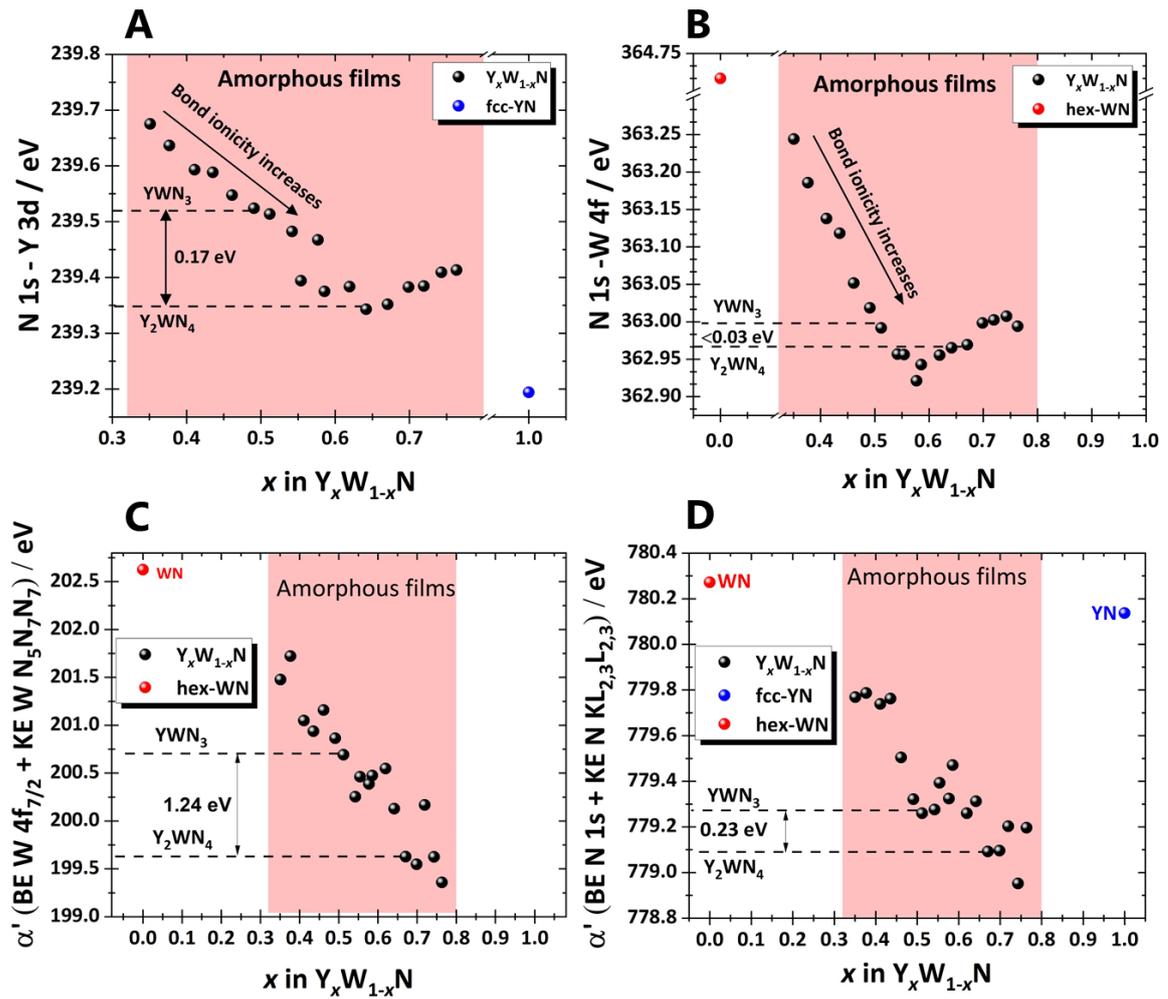

**Fig. 3 Chemical environment analysis.** Binding energy difference between the N 1s - Y 3d (A) and the N1s - W4f (B) core levels as a function of composition for $Y_xW_{1-x}N$ thin film grown at $T_s$ = 100 °C. Auger parameter of (C) W and (D) N plotted as a function of composition of $Y_xW_{1-x}N$ thin films grown at $T_s$ = 100°C.

Composition analysis of both $YWN_3$ and $Y_2WN_4$ samples using Rutherford backscattering (RBS) analysis as well as elastic recoil detection analysis (ERDA) reveals an almost perfect stoichiometry (Supplementary Table 2). The metal ratio is W: 18.7 at.% and Y: 19.5 at.%, whereas the nitrogen content is slightly higher with N: 58 at.% with marginal O impurity of 2.0 at.% for $YWN_3$. A decrease in the N-to-metal ratio with increasing of $T_s$ is associated to recombination and desorption of N at a high deposition temperature [24]. Only slight deviation from stoichiometry is observed for $Y_2WN_4$ grown at $T_s$ = 400°C and a marginal O impurity.

The local chemical environment is probed *via* ultra-high vacuum (UHV) transfer X-ray photo electron spectroscopy (XPS) by measuring W 4f, Y 3d, and N 1s photoemission lines and W $N_5N_7N_7$ and N $KL_{2,3}L_{2,3}$ Auger lines to calculate the N-to-metal core-level binding energy difference and modified Auger parameter (AP), respectively. The latter two XPS-derived parameters are insensitive to charging effects and erroneous binding energy calibration and used here due the insulating properties of the films [25]. Charge transfer analysis (Fig. 3A, B) determined here as a difference in peak position between N1s core level and metal core-levels (Y 3d and W 4f) reveals an increase in charge transfer from the metals to N as the Y-concentration $x$ increases. This suggests a decrease in covalence, hence a more ionic nature of the bonds, in Y-rich films. This is reflected in the higher resistivity of the Y-rich compound: $2.5 \times 10^8$ Ohm·meter for $Y_2WN_4$ and $5.0 \times 10^5$ Ohm·meter for $YWN_3$ films grown at $T_s = 400$ °C. This is also reflected in the N and W AP evidenced by a shift to lower values as a function of $x$, Fig. 3C, D. This shift is typically indicative of changes in the coordination number as well as the distance and electronic polarizability of the core-ionized atom's nearest-neighbors [26]. Considering that the oxidation state of W is 6+ in both $YWN_3$ to $Y_2WN_4$ thin films, the change of W AP by 1.24 eV (Fig. 3C) can be assigned mostly to a decreased covalence of the nearest-neighbor bonds as films become Y-rich, which is in line with the much lower electronegativity of Y compared to W. Only a 0.23 eV difference in N AP for $YWN_3$ and $Y_2WN_4$ (Fig. 3D) indicates a consistent oxidation state of N with minor effects related to changes in the nearest-neighbor electronegativity. Strikingly the charge transfer analysis reveals a minimum around the most chemically stable stoichiometries $YWN_3$ and $Y_2WN_4$ (see Fig. 3). For W-rich films, the N 1s – W 4f and AP values continuously shift toward the values recorded for the crystalline WN phase. This suggests the increase in the driving force to phase separation via precipitation of WN.

## Functional properties of YWN$_3$ and Y$_2$WN$_4$

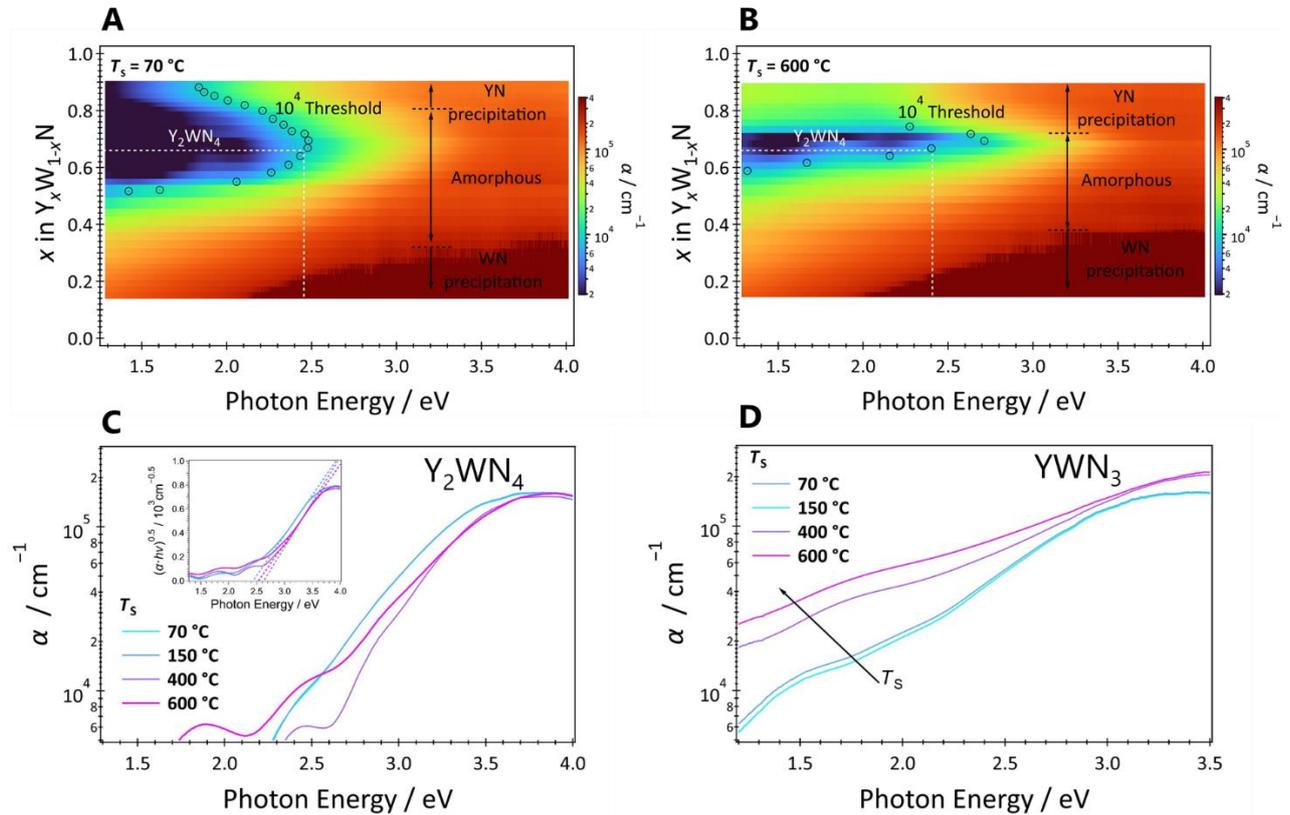

**Fig. 4 Optical properties of Y$_2$WN$_4$ and YWN$_3$ thin films.** Absorption coefficient from UV−Vis mapping for W-Y-N thin films synthesized at (A) $T_s$ = 70 °C and (B) $T_s$ = 600 °C. The $10^4$/cm threshold is displayed as circles. All thin films are grown on borosilicate glass substrates. Absorption coefficient for the (C) Y$_2$WN$_4$ and (D) YWN$_3$ films grown at different $T_s$ with the Tauc plot for an indirect transition as an inset in (C) for films grown at different $T_s$.

Considering the amorphous structure of YWN$_3$ and Y$_2$WN$_4$ films but distinct constituent chemical bonding we attempt to determine functional properties of these two different amorphous compounds. Using an automated combinatorial UV−Vis platform [27], we investigated the effect of the substrate temperature on the optical properties of our Y-W-N libraries. Films grown at $T_s$ = 70 °C (Fig. 4A) exhibit decreased overall optical absorption compared to those grown at $T_s$ = 600 °C (Fig. 4B). As previously reported by our group [27,28], this may be related to decreased optical scattering otherwise resulting from the precipitation of another phase. Furthermore, the sharpest absorption onset was observed for Y$_2$WN$_4$ as evident by the continuously low sub-bandgap absorption across all investigated deposition temperatures (Fig. 4C). From Tauc plot analysis, an indirect band gap of ~2.5 eV was determined for Y$_2$WN$_4$. Complementary ellipsometry measurements (Fig. S8) further specified this value to 2.24 eV, which lies well in agreement with typical errors associated with Tauc-plot analysis [29]. In contrast, YWN$_3$ exhibited a lowered steepness of the absorption onset with higher substrate temperatures, in agreement with a higher driving force for precipitation at elevated temperatures. The decreased real (Fig. S8A) and imaginary dielectric constants (Fig. S8B) especially throughout the measured spectrum for YWN$_3$ validate these observations.

We also perform mechanical properties mapping of the amorphous phase space using an automated nano-indentation system to study mechanical robustness of the films grown at $T_s$ = 400, 289, and 215 °C (Supplementary Figure S9). All films show a negligible dependence from $T_s$ within measurement error suggesting a good thermal stability of the amorphous phase. However, the hardness of YWN$_3$ thin films is only slightly higher than that of Y$_2$WN$_4$ at all $T_s$, in particular 8.8 ± 0.2 GPa and 8.2 ± 0.2 GPa, respectively, at $T_s$ = 215 °C. Importantly, this difference can be assigned to increased covalence of the nearest-neighbor bonds revealed by XPS analysis for YWN$_3$ as well as possible segregation of WN in this compound.

For further assessment of potential functional properties of these compounds we investigate the oxidation resistance, thermal stability and Cu diffusion barrier performance of Y$_2$WN$_4$ and YWN$_3$ thin films grown at $T_s$=400 °C by performing *in situ* temperature- and time-dependent XRD measurements (see Materials and Methods). Y$_2$WN$_4$ films exhibit much better thermal stability than YWN$_3$ films (see Fig. 5A and Fig. 5B). This is evidenced as a steep reduction of the broad-peak intensity after 50 min at 700 °C for Y$_2$WN$_4$ (Fig. 5A and Fig. S10A) whereas YWN$_3$ films demonstrate systematic reduction of the intensity already after 200 °C (Fig. 5B and Fig. S10A). The diffusion barrier performance was evaluated by depositing 10 nm Y$_2$WN$_4$ and YWN$_3$ barrier layers on Si wafers followed by a 100 nm thick Cu film. We compare their performance with 10 nm thick TiN thin films, which are widely recognized for their effectiveness as diffusion barriers in microelectronic devices [30]. The layer stacks are heated *in situ* while monitoring potential copper silicide formation. Remarkably, both amorphous nitrides exhibit comparable Cu diffusion barrier performance and both outperform TiN diffusion barrier of the same thickness (Fig. 5C-D and Fig. S11). The Cu$_3$Si XRD peak, which indicates Cu diffusion and subsequent silicide formation can be detected only for TiN barrier layer, Fig. S11. TiN diffusion barrier failure temperature is as high as 600 °C whereas Y$_2$WN$_4$ withstands up to 800 °C as the signal from the topmost Cu film can be detected even after 150 min at 800 °C without noticeable formation of Cu$_3$Si or other Si containing phases. At the same conditions, YWN$_3$ barrier layer decomposes leading to the formation of *bcc*-W phase already after 40 min at 700 °C. In terms of diffusion barrier failure temperature Y$_2$WN$_4$ outperforms or is comparable to other W containing amorphous nitride compounds [31–33]. Y$_2$WN$_4$ films show also a better oxidation resistance as the onset of oxidation starts after 30 minutes holding at 700 °C while YWN$_3$ oxidizes starting from 10 minutes at 700 °C (Fig. S10C, Fig. S12).

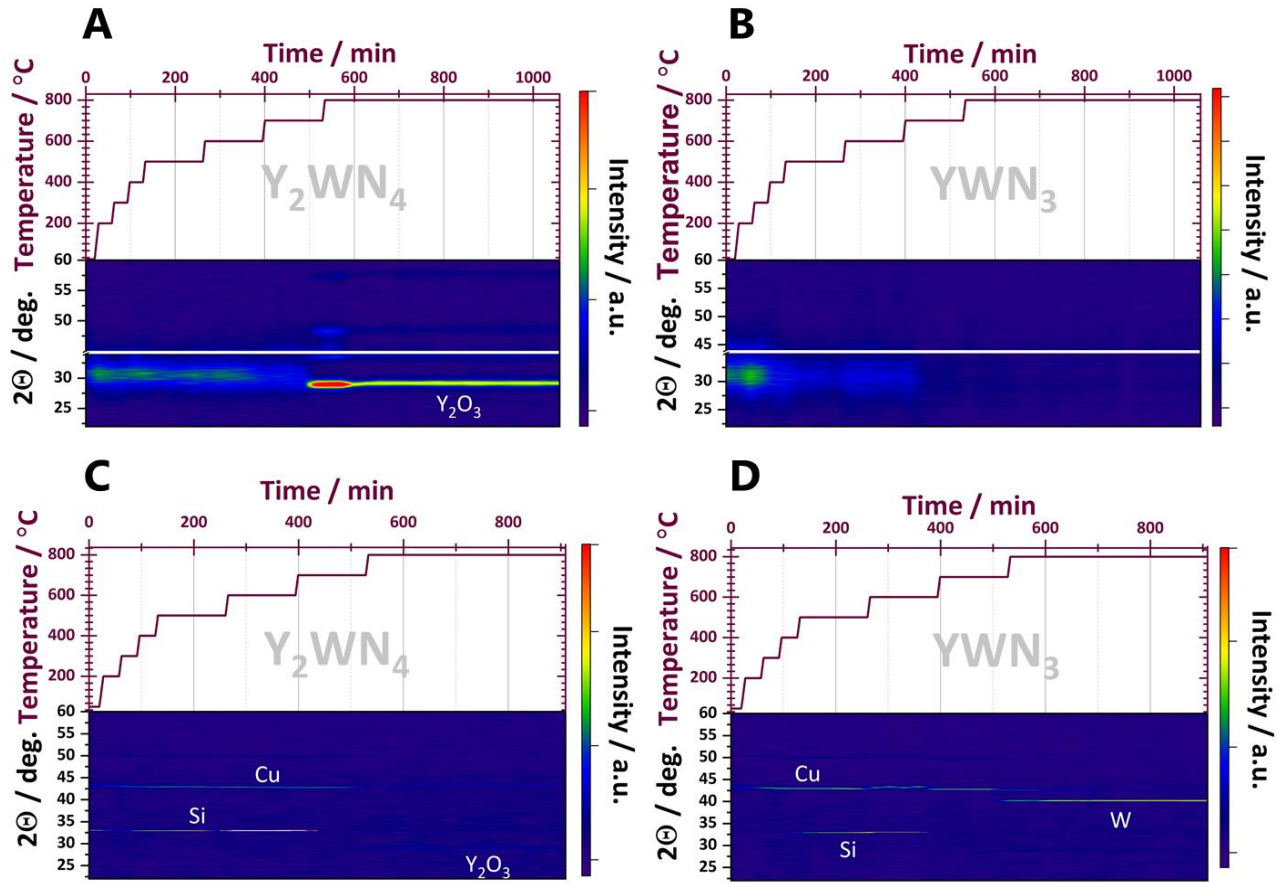

**Fig. 5 Functional properties of Y$_2$WN$_4$ and YWN$_3$.** *In situ* temperature- and time-dependent XRD results performed in the atmosphere of N$_2$+5% H$_2$ for (A) Y$_2$WN$_4$ and (B) YWN$_3$ thin film grown on sapphire substrates. *In situ* temperature- and time-dependent XRD results performed in the atmosphere of N$_2$+5% H$_2$ for a 10 nm-thick (C) Y$_2$WN$_4$ and (D) YWN$_3$ thin film grown on Si substrates and covered with a 100 nm-thick Cu thin film. Despite the experiments are performed in an inert atmosphere, Y$_2$O$_3$ forms in Y-rich samples due to a very strong oxygen affinity of Y.

## Discussion

Y$_2$WN$_4$ and YWN$_3$ thin films grown at $70 \leq T_s \leq 600$ °C are amorphous confirming the computational random structure sampling results. XRD patterns for all predicted space groups for YWN$_3$ available in the Materials Project database presented in Fig. 1 D shows only crystalline phases for the given stoichiometry whereas our calculated XRD patterns for both Y$_2$WN$_4$ and YWN$_3$ (Fig. 1C, D) reveal amorphous phase formation for both compounds that well agrees with experiment. Moreover, theoretical calculations show that Y$_2$WN$_3$ is less prone to crystallization than YWN$_3$. This is confirmed experimentally allowing the former to retain its amorphous structure and associated oxidation resistance and barrier performance against Cu diffusion at high temperatures during *in situ* XRD tests. This remarkable temperature stability and diffusion barrier performance is most probably associated with the ionic nature of the chemical bonding in this Y-rich compound revealed by XPS. This brings the $Y_2WN_4$ compound in line with other application-relevant ionic amorphous materials, like amorphous aluminum oxide, silicon nitride and silicon oxide. Therefore, bond iconicity can be considered as a criterion for the design of stable amorphous compounds for high temperature applications. Additionally, $Y_2WN_4$ and $YWN_3$ demonstrate dissimilarities in optical and mechanical properties. $YWN_3$ shows lower steepness of the absorption onset at high deposition temperature that is typical for precipitation. However, no sign of crystalline precipitates is revealed by the HRSTEM analysis of $YWN_3$ samples grown at $T_s$ = 400 °C as well as no signs of W segregation. However, as the films grown at $T_s$ = 100 °C become W-rich, the W and N APs shift towards values typical of WN. This demonstrates a strong driving force for, and likelihood of, WN precipitation in $YWN_3$ films. This aligns with the slightly higher hardness of $YWN_3$ films compared to $Y_2WN_4$ films, due to the covalent and thus stronger nature of W-N bonding.

## Conclusions

Our work demonstrates a computationally guided approach to the discovery of novel amorphous materials. Through computational structure sampling, we identified two hard-to-crystallize compounds, $Y_2WN_4$ and $YWN_3$, within the YW-WN pseudo-binary phase space. These computational predictions were successfully verified by synthesizing the corresponding thin films. Additionally, by depositing combinatorial thin film libraries, we determined a broad composition and deposition temperature range where Y-W-N thin films exhibit an amorphous structure. Tuning the chemical composition across this range while maintaining the amorphous structure allows for customization of the structural and functional properties of these compounds. Notably, we discovered an amorphous $Y_2WN_4$ compound demonstrating a distinct band gap, remarkable thermal stability and excellent performance as a diffusion barrier against Cu diffusion. The approach presented here enables a pathway for the discovery of entirely new amorphous materials with unique functionalities.

## Acknowledgements

O.V.P. acknowledges funding from the SNSF (grant no. 227945). S.Z. acknowledges research funding from Empa Internal Research Call 2020. J. P. acknowledges funding by the SNSF (grant no. 196980). A.W. acknowledges funding from the the Strategic Focus Area–Advanced Manufacturing (SFA–AM) through the project Advancing manufacturability of hybrid organic–inorganic semiconductors for large area optoelectronics (AMYS). The computational work is supported by the United States National Science Foundation, Grant No. DMR-1945010 and was performed using computational resources provided by Colorado School of Mines. The authors thank Arnold Müller and Christof Vockenhuber from Laboratory of Ion Beam Physics at ETH Zurich for RBS and ERDA analysis. The authors further acknowledge Erwin Hack for help with the ellipsometry analysis, Kerstin Thorwarth for help with the Y-W-N depositions and Stefanie Frick for helpful discussions and feedback on the manuscript draft.

The author contributions are as follows: S.Z., O.V.P., V.S. and S.S. conceived the study. O.V. P. and S. Z. performed the Y-W-N deposition and materials characterization incl. XRF, XRD and XPS mapping supervised by S. S. J.P. supported HiPIMS depositions (not reported) and performed EDX analysis. A.W. performed the optical characterization. A. S. performed the microstructural characterization supervised by J. M., C. C. performed the high-temperature XRD measurements. All computational work was performed by V. S. O.V.P. wrote the original draft of the manuscript with contributions from V.S., S. S. and A. W. and A. S. All authors reviewed, commented on and approved the final version of the manuscript.

**Supplementary information for**

# Computationally-guided discovery and synthesis of the amorphous nitride $Y_2WN_4$

O.V. Pshyk[a], S. Zhuk[a], J. Patidar[a], A. Wieczorek[a], A. Sharma[b], J. Michler[b], C. Cancellieri[c], V. Stevanovic[d,*], S. Siol[a,*]

a) Empa – Swiss Federal Laboratories for Materials Science and Technology,
   Laboratory for Surface Science and Coating Technologies, Dübendorf 8600, Switzerland
b) Empa — Swiss Federal Laboratories for Materials Science and Technology,
   Laboratory for Mechanics of Materials and Nanostructures, Thun 3602, Switzerland
c) Empa – Swiss Federal Laboratories for Materials Science and Technology,
   Laboratory for Joining Technologies and Corrosion, Dübendorf 8600, Switzerland
d) Colorado School of Mines, Golden, 80401, CO, United States

*Corresponding authors: vstevano@mines.edu, sebastian.siol@empa.ch*

**Materials and Methods**

*Computational structure sampling.*

The first principles random structure sampling proceeds as follows. For a given chemical composition and predefined number of atoms (4 formula units in this case), a large number of periodic structures with random lattice parameters (a, b, c, α, β, γ) and random atomic positions are generated. These are then relaxed to the closes local minimum on the PES using density functional theory (DFT) or related total energy method, and some gradient based relaxation methods (conjugate gradient, steepest decent etc.). Afterwards, the relaxed structures are grouped into classes of equivalent structures. The number of relaxed structures in each class, or its frequency of occurrence, can then be used as a measure of the size of the corresponding PES local minimum or, more precisely, the volume of configuration space occupied by a given local minimum. In our previous works we have shown that known crystalline phases across many different chemistries consistently appear as high-frequency structures in random sampling [16–18,34] and that structural characteristics of amorphous phases (amorphous Si) can be accurately reproduced using the entire ensemble of relaxed random structures as a representation of an amorphous phase [35]. Herein, we advance these concepts and show that the first-principles random structure sampling also provides insights into the stability of amorphous phases toward crystallization and can be used to identify chemistries that are hard to crystalize.

The specific version of the first-principles random structure sampling adopted here involves modifications of the random structure generations designed specifically for partially ionic systems to bias the sampling toward structures that are dominantly cation-anion coordinated. This is done by discretizing the space inside the unit cell and distributing atoms over the two interpenetrating grids of points. Cations are then distributed randomly over the cation grid and anions over the anion grid. To ensure that no two atoms are too close to each other (or on top of each other) a gaussian of a certain width is placed on every grid point occupied by an atom. The next atom is then placed on a point randomly chosen from a set of grid points that have the value of the sum of the gaussians below a certain threshold. Structure relaxations are performed using the standard first-principles setup designed to accurately reproduce compound enthalpies of formation [36]. The electron-electron interactions are treated using the PBE exchange-correlation functional [37] with the addition of the rotationally invariant Hubbard U term [38]. In accordance with Ref. [36] the uniform U=3 eV is used for both Y and W. It should be noted that these (thermochemical) U values are appropriate for total energy and formation enthalpy calculations and are by no means meant to produce accurate electronic structure. The projector augmented wave (PAW) method [39] as implemented in the VASP computer code [40]. Plane-wave cutoff of 340 eV is used in all calculations, which is ~20 % higher than the recommended value by the used pseudopotentials (Y_sv and W_pv, N_s). Automatic generation of the Γ-centered k-point grid is employed with the Rk value of 20.

A total of 2,000 random structures with 4 formula units for each Y1W1N3 and Y2W1N4 have been generated and relaxed. We showed in our previous works that an ensemble of structures with 20 or more atoms is sufficient to provide enough diversity for the structural features of amorphous Si to be well described [40]. Also, 2,000 structures are more than enough to obtain converged structure factor. The subsequent classification of structures is performed based on 4 criteria: (a) the space group assignment is the same between two structures, (b) total energy within 10 meV/atom, (c) volumes with 5 %, and (d) coordination numbers up to 3rd shell are the same. The structural (and other) properties of the amorphous phase are calculated though the ensemble averaging of the properties evaluated for each individual structure.

*Thin-film growth*

Y-W-N thin films are grown by reactive radio-frequency (RF) magnetron co-sputtering of Y and W targets (50.8 in diameter, 99.99 % purity). Deposition of thin films is performed in an AJA 1500-F sputtering system pumped down to a base pressure lower than $10^{-6}$ Pa. The type-2 unbalanced magnetrons are operated in a closed-field magnetic configuration with the reactive N gas routed directly into the chimneys of the magnetrons. The deposition is carried out in a confocal sputter-up geometry with oblique deposition angles to synthesize combinatorial sample libraries with compositional gradients. Temperature gradients in the direction orthogonal to the composition gradient was achieved by partial clamping the substrate to the heated holder. Both magnetrons operate in a closed-field magnetic configuration. The deposition is carried out in a mixed atmosphere of $N_2$ (7 sccm) and Ar (10 sccm) at a pressure of 0.33 Pa. The thickness of the thin film libraries varied between 350 nm and 450 nm depending on composition. The thin film library for nano-indentation mapping is grown to a thickness of 1.0 to 1.3 µm. The growth of the selected films ($YWN_3$ and $Y_2WN_4$) is performed on rotating substrates. The films for oxidation resistance and thermal stability investigation are growth to the thickness of 500 nm. The films for resistance measurements with a thickness of 110 nm are deposited on a Ti/Pt bottom electrode, and small platinum (Pt) pads were deposited over the film to function as the top electrode. 1.1 mm thick borosilicate glass (EXG) substrates of 50.8 mm × 50.8 mm size are used for combinatorial thin film growth while single-crystalline (001) Si and (0001) α-$Al_2O_3$ substrates are used for the deposition of the serial depositions. The substrates were ultrasonically cleaned in acetone and ethanol before the deposition. TiN and Cu thin films for the diffusion barrier performance study are grown by direct current (DC) magnetron sputtering of Ti and Cu tar-gets, respectively (50.8 in diameter, 99.99 % purity) at a substrate temperature of 400°C. The deposition of TiN is carried out in a mixed atmosphere of $N_2$ (8 sccm) and Ar (18 sccm) at a pressure of 0.5 Pa. A radio-frequency (RF, 13.56 MHz) substrate bias of -75 V is applied during the growth of TiN thin films. To remove surface oxides and contaminants from the Si substrates before the deposition of a barrier layer, RF substrate cleaning is performed by applying a power of 6 W for 5 minutes.

*Structural and chemical composition characterization.*

Structural characterization of serial thin films and mapping of combinatorial libraries is performed in a Bruker D8 X-ray diffraction (XRD) system with Cu Kα X-ray operating in Bragg−Brentano geometry. A Fischer XDV-SDD X-ray fluorescence (XRF) system equipped with a Rh X-ray source is used for W to Y ratio determination and thickness measurements. The latter measurements are preliminary calibrated using 13 MeV 127-I Elastic Recoil Detection analysis (ERDA) and Rutherford backscattering spectrometry (RBS). XPS measurements are performed using a Physical Electronics Quantera II Hybrid spectrometer using monochromated Al-Kα radiation. The base pressure during the spectra acquisition is below $10^{-6}$ Pa. Charge neutralization is achieved using a dual beam charge neutralization system based on a low-energy electron flood gun and a low energy positive ion source. The binding energy scale is calibrated by examining sputter-cleaned Au, Ag, and Cu reference samples according to the recommended ISO standards for monochromatic Al-Kα sources that place Au $4f_{7/2}$, Ag $3d_{5/2}$, and Cu $2p_{3/2}$ peaks at 83.96, 368.21, and 932.62 eV respectively [41,42]. The samples are transferred from the deposition chamber to the spectrometer using a UHV transfer chamber. The samples are stored at a pressure below $10^{-4}$ Pa throughout the transport process.

TEM, SAED and STEM of the films is performed in a Themis 200 G3 spherical aberration (probe)-corrected TEM using an accelerating voltage of 200 keV. The cross- sectional TEM samples are prepared by using the lift-out method in a dual-beam focussed ion beam (FIB-SEM Tescan Lyra FEG system). The final TEM sample thinning was done at an ion current of 24 pA at 5 kV to a thickness of 80 nm.

*Functional properties.*

A home-built automated UV−Vis−NIR mapping spectrophotometer equipped with deuterium UV and tungsten−halogen broad-band light sources as well as CCD spectrometers from Ocean Insight is used for transmittance and reflectance spectra measurements and further calculation of the absorption coefficient.

Nanoindentation measurements are performed using a Picodentor HM500 equipped with a Berkovich diamond tip. The indentation depth is limited to less than 15 % of the total film thickness to minimize the elastic behavior of the substrate during nanoindentation. 25 indents are performed for each sample. Hardness was determined following Oliver and Pharr method [43].

*In situ* XRD analyses of the samples upon annealing were conducted on the PANalytical X'Pert-Pro diffractometer system equipped with the CuK$\alpha_{1,2}$ radiation at 40 kV/40 mA and Anton Paar XRK 900 furnace operated in a continuous flow of selected gas or in air. The θ–2θ scans are recorded in the Bragg–Brentano geometry in the range between 20-60°, with a step of 0.026° with the sample continuously heated up from room temperature to 800 °C and afterward cooled down to room temperature. In the present work, for diffusion barrier performance and thermal stability study the measurements are performed in the $N_2$ flow with 5% $H_2$. Oxidation resistance study is performed in ambient air.

Spectroscopic ellipsometry is performed using a J.A. Woollam M2000-VI at three incidence angles (50°, 60°, 70°). The layers were fitted with a Cauchy model, including an Urbach absorption term. Alternatively, a fit with Lorentz oscillators is equivalent. To obtain the exact position of the absorption band, we replace the Cauchy model with Urbach absorption by a Lorentz Oscillator and refit.

**Supplementary notes**

Supplementary note 1

The most basic approach for selecting potential candidates for amorphous nitride thin films is based on the consideration of a possible miscibility between the ground state binary phases or other possible polymorphs existing for a given compound system. Considering the latter, the systems with dissimilar ground state structures are likely form amorphous phase upon alloying because of a high mixing enthalpy and low preference for the formation of solid solution. Although the ground state structure within the W-N compound system is corundum-derived $W_2N_3$ crystallizing in the monoclinic Cm space group (Fig. S1), W-N thin films crystalize in a face-centered cubic NbO structure β-$W_2N$ containing a regular array of vacancies on the cation and anion sub-lattices or a hexagonal δ-WN at a high $N_2$ flow rate under highly nonequilibrium physical vapour deposition (PVD) conditions [24,44]. At the same time YN thin films crystallize in a NaCl-structured phase. This implies that $Y_xW_{1-x}N$ pseudo-binary system is the potential candidate with a preference to form an amorphous phase.

Supplementary note 2

WN thin films have a relatively low oxidation temperature onset of 600 °C leading to thin film material sublimation due to the formation of a volatile $WO_x$ [44]. In contrast, oxidation temperature onset of YN is even lower [45], most probably around room temperature as the oxidation of YN thin films takes place even under UHV conditions [45,46]. Therefore, it is difficult to utilize this materials as oxidation resistant and protective coating. However, we propose that alloying of WN and YN may allow to combine benefits of the two compounds while mitigating the disadvantages. First of all, given by a potential earlier oxidation of YN, it may form a passivation layer in analogy to $Al_2O_3$ [47] protecting the entire film from further oxidation while amorphous structure of Y-W-N films can potentially postpone/reduce diffusion of species thus limiting the oxidation process. Motivated by the fact that the Y-W-N pseudo-binary system has not been studied, to the best of our knowledge, the phenomena arising due to a mixture of two dissimilar nitride compounds may be of outstanding interest from the scientific point of view as well as from the application potential.

**Supplementary Figures**

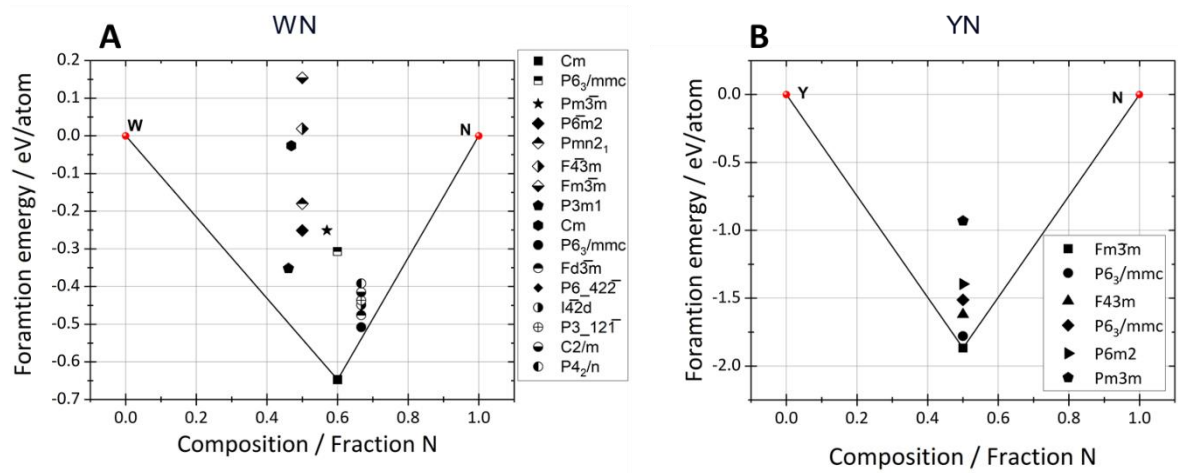

**Fig. S1** Phase diagrams for (A) $W_{1-x}N_x$ and (B) $Y_{1-x}N_x$ systems [13].

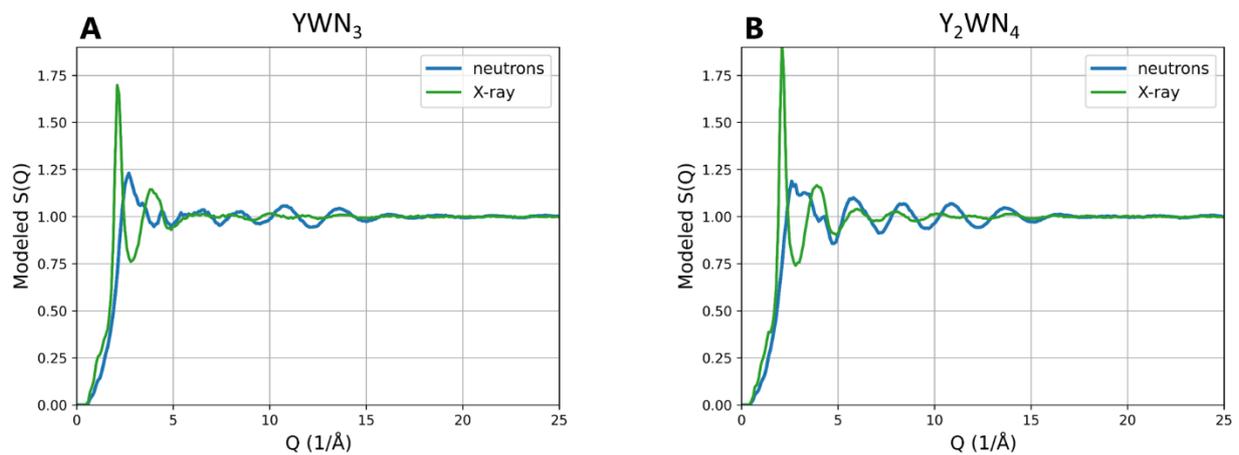

**Fig. S2** Modeled neutron and X-ray structure factors averaged over all random structures representing the amorphous phase for (A) $YWN_3$ and (B) $Y_2WN_4$.

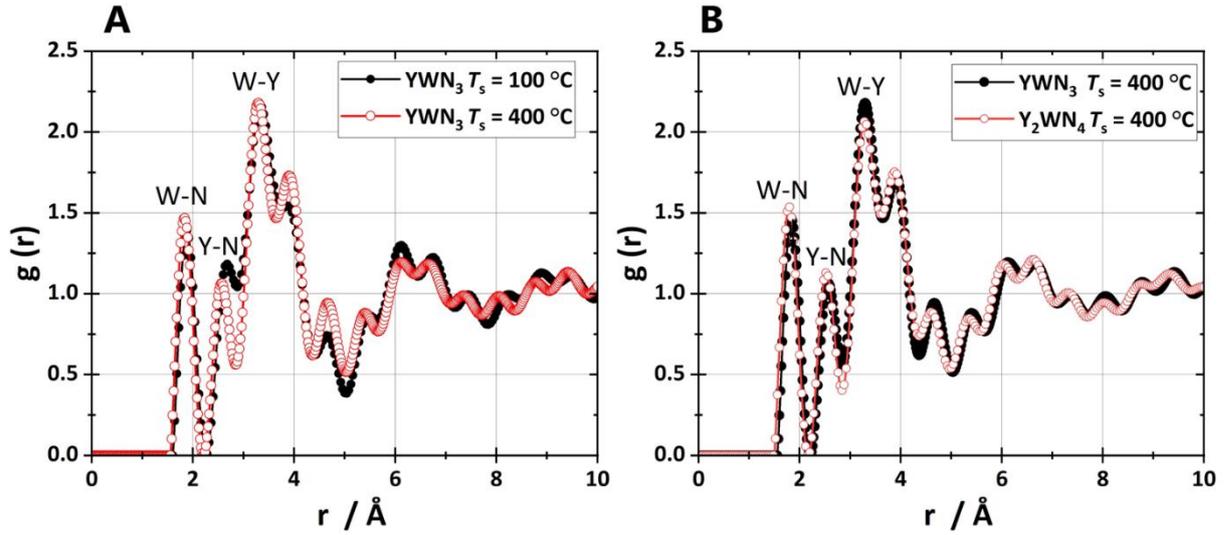

**Fig. S3** SAED-derived pair distribution function (PDF) for (A) YWN$_3$ grown at $T_s$ = 100 and $T_s$ = 400 °C and for (B) YWN$_3$ and Y$_2$WN$_4$ films grown at $T_s$ = 400 °C.

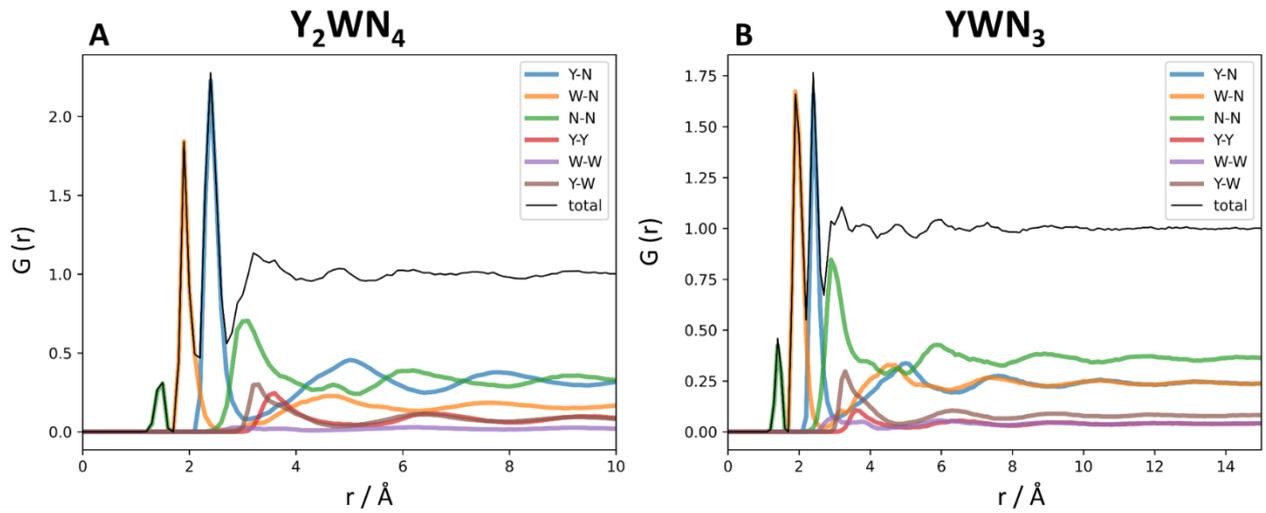

**Fig. S4.** Theoretical total and partial geometrical PDFs for Y$_2$WN$_4$ and YWN$_3$ compounds.

**Supplementary Table 1**. Comparison of bond lengths constituting YWN$_3$ and Y$_2$WN$_4$ compounds determined computationally and experimentally from SAED-derived PDF analysis. The coordination number is given in brackets.

|  |  | $T_s$ / °C | Bond length / Å | | | | |
|---|---|---|---|---|---|---|---|
|  |  |  | **N-N** | **W-N** | **Y-N** | **W-Y** | **Y-W** |
| **Experimental** | **YWN$_3$** | 100 | – | 1.9 | 2.7 | 3.3 | 3.3 |
|  |  | 400 | – | 1.8 | 2.6 | 3.3 | 3.3 |
|  | **Y$_2$WN$_4$** | 400 | – | 1.8 | 2.5 | 3.3 | 3.3 |
| **Calculated** | **YWN$_3$** | – | 1.4 (0.24) | 1.9 (4.73) | 2.4 (6.24) | 3.3 (6.64) | 3.3 (6.64) |
|  | **Y$_2$WN$_4$** | – | 1.5 (0.18) | 1.9 (4.40) | 2.4 (5.77) | 3.3 (8.62) | 3.3 (4.31) |

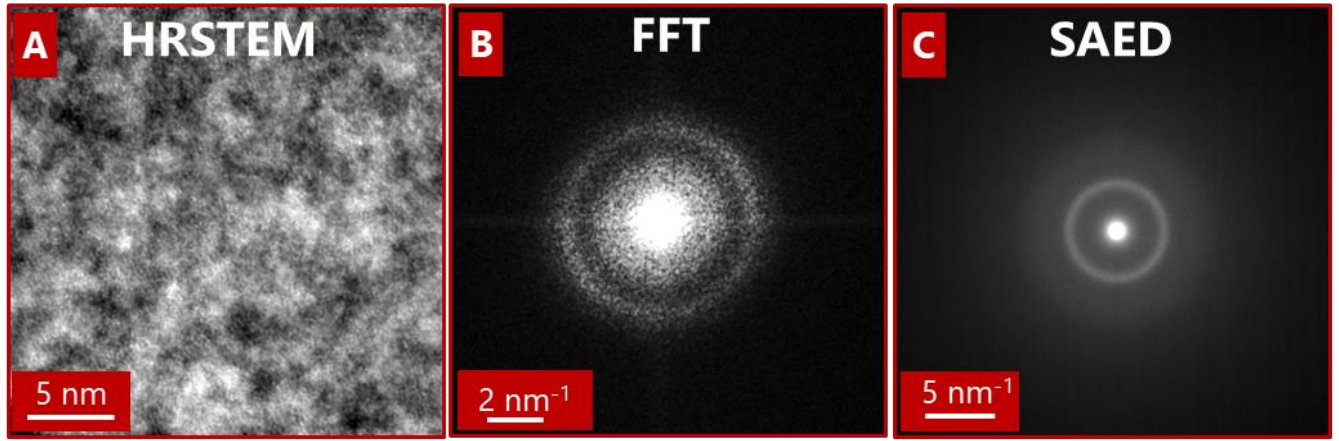
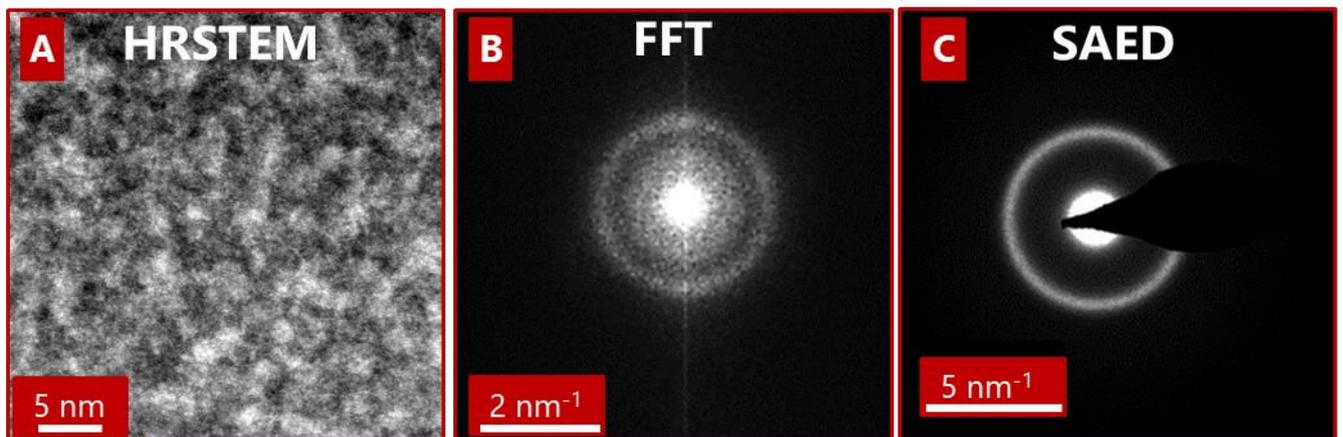
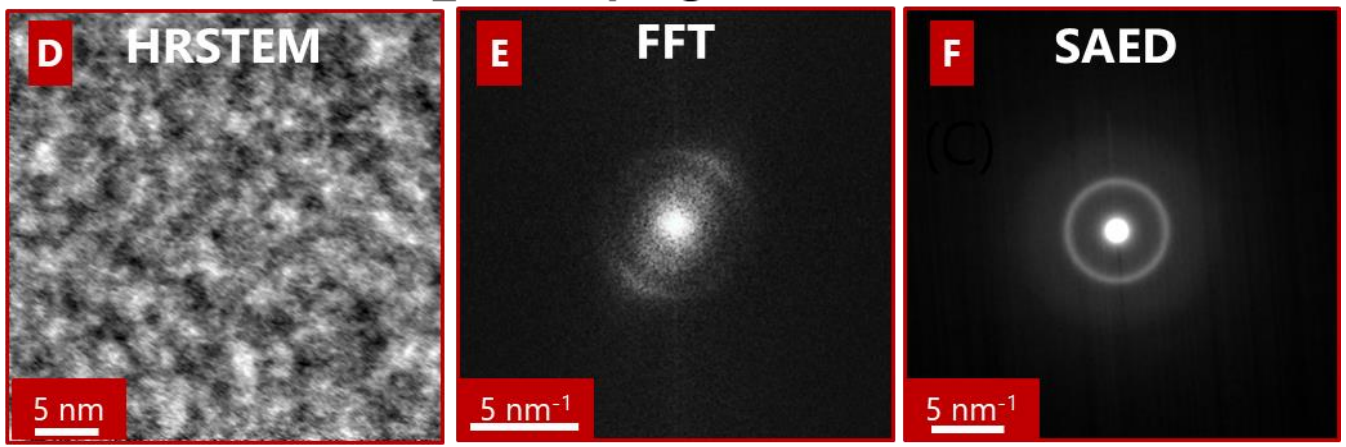

**Fig. S5** HR-STEM images, corresponding FFT patterns, and SAED patterns for (A-D) YWN$_3$ films grown at $T_s$ = 100 °C, (B-F) YWN$_3$ films grown at $T_s$ = 400 °C and (G-I) Y$_2$WN$_4$ films grown at $T_s$ = 400 °C. All thin films are grown on borosilicate glass substrates.

Pshyk *et al.* Empa 2024 26

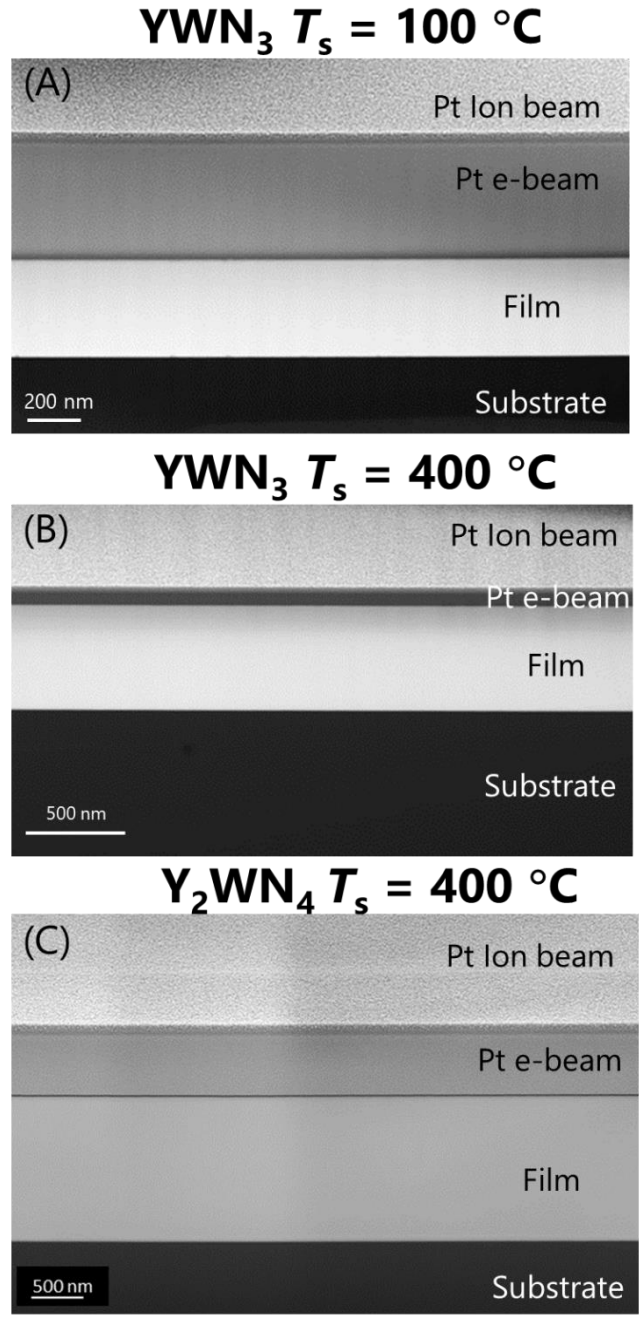

**Fig. S6** Overview HAADF-STEM images for (A) YWN$_3$ films grown at $T_s$ = 100 °C, (B) YWN$_3$ films and (C) Y$_2$WN$_4$ films grown at T$_s$ = 400 °C. All thin films are grown on borosilicate glass substrates.

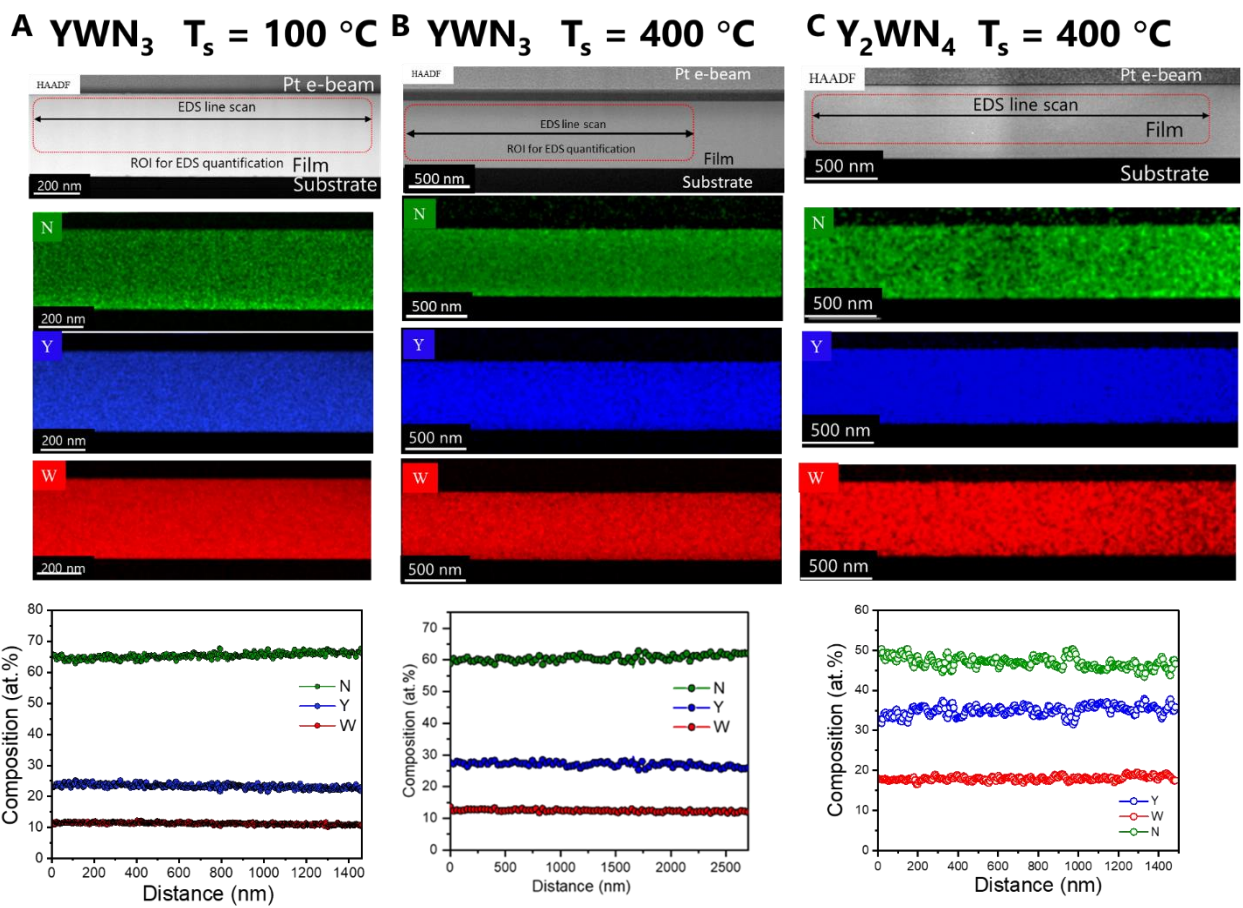

**Fig. S7** Overview HAADF-STEM images and corresponding EDS elemental maps for N, Y, and W and EDS linescans for (A) YWN$_3$ films grown at $T_s$ = 100 °C, (B) YWN$_3$ films and (C) Y$_2$WN$_4$ films grown at $T_s$ = 400 °C. All thin films are grown on borosilicate glass substrates.

**Supplementary Table 2**. Chemical composition of YWN$_3$ films grown at $T_s$ = 100 °C, YWN$_3$ films grown at $T_s$ = 400 °C and Y$_2$WN$_4$ films grown at $T_s$ = 400 °C measured by means of RBS/ERDA. All thin films are grown on borosilicate glass substrate.

| Chemical formulae | $T_s$ / °C | Composition / at.% | | | | | | |
|---|---|---|---|---|---|---|---|---|
| | | Y | W | N | O | Ar | C | H |
| YWN$_3$ | 100 | 18.7 | 19.5 | 58.0 | 2.0 | - | 0.6 | 0.54 |
| | 400 | 22.1 | 18.4 | 56.4 | 2.1 | 1.0 | - | - |
| Y$_2$WN$_4$ | 400 | 28.0 | 13.8 | 53.8 | 3.5 | 0.9 | - | - |

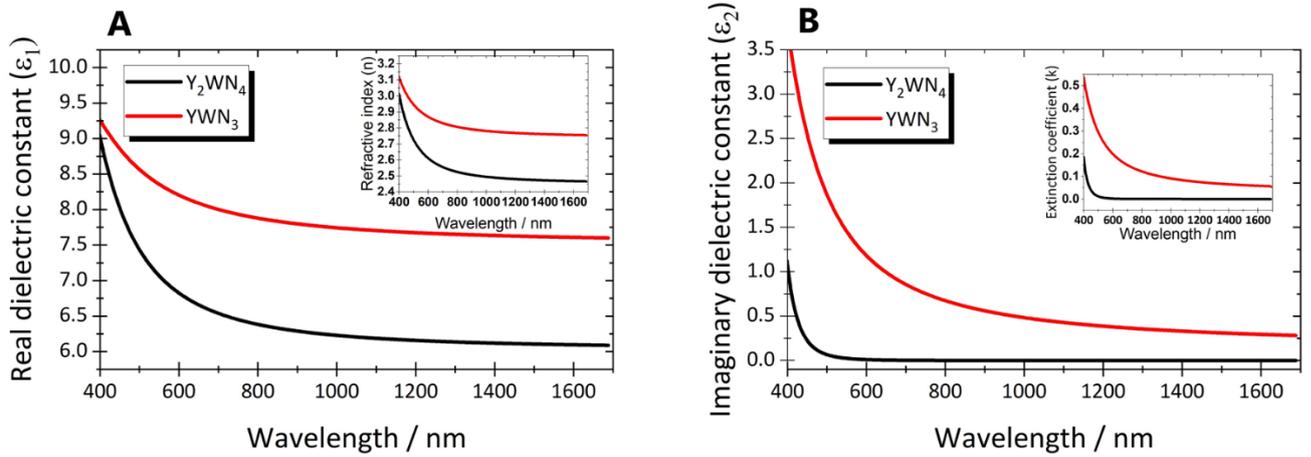

**Fig. S8.** (A) real ($\varepsilon_1$) and (B) imaginary ($\varepsilon_2$) dielectric constant for YWN$_3$ and Y$_2$WN$_4$ films grown at $T_s$ = 400 °C measured by spectroscopic ellipsometry with refractive index and extinction coefficient as insets in (A) and (B), respectively. All thin films for ellipsometry measurements are grown on Si substrates.

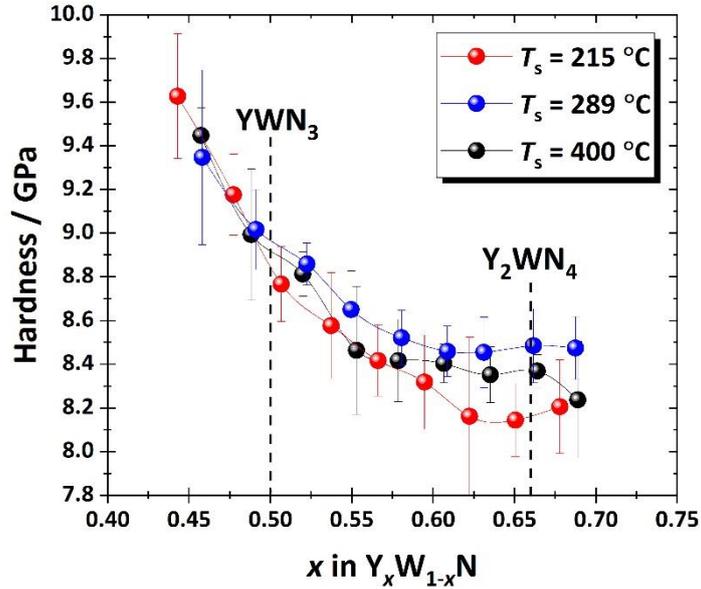

**Fig. S9.** Nanoindentation hardness for $Y_{1-x}W_xN$ thin film libraries grown on a borosilicate glass substrate at temperature of 215, 289, and 400 °C.

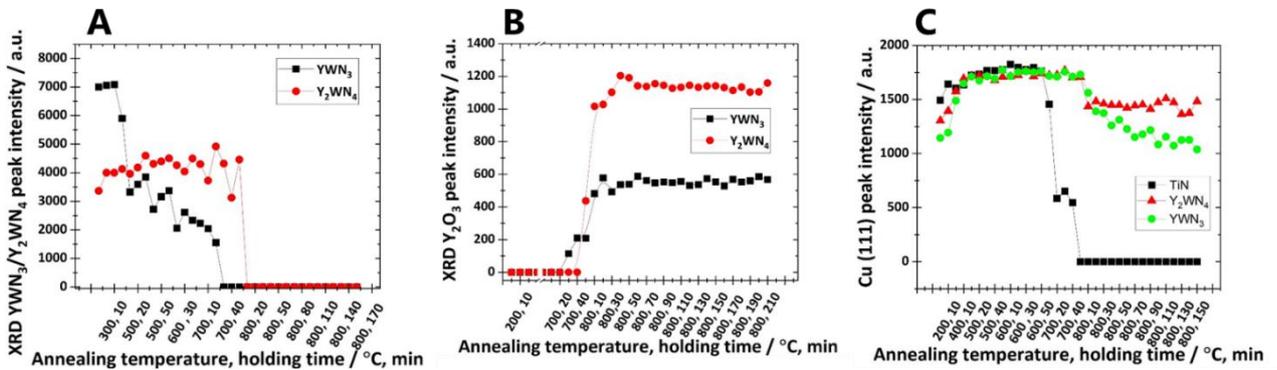

**Fig. S10.** (A) *in situ* temperature-dependent XRD intensities from a broad diffraction reflection plotted as a function of temperature and holding time for $YWN_3$ and $Y_2WN_4$ thin films annealed in $N_2$+ 5% $H_2$ atmosphere. (B) *in situ* XRD peak intensity of $Y_2O_3$ oxide phase plotted as a function of temperature and holding time from $YWN_3$ and $Y_2WN_4$ thin films on sapphire substrates annealed in ambient atmosphere. (C) Summary plots from in situ XRD of 10 nm $YWN_3$ and $Y_2WN_4$ thin films covered with 100 nm Cu thin films (Cu seed stack layer) annealed in $N_2$+$H_2$ atmosphere showing measured XRD peak intensities from Cu (111) diffraction reflection as a function of temperature an holding time.

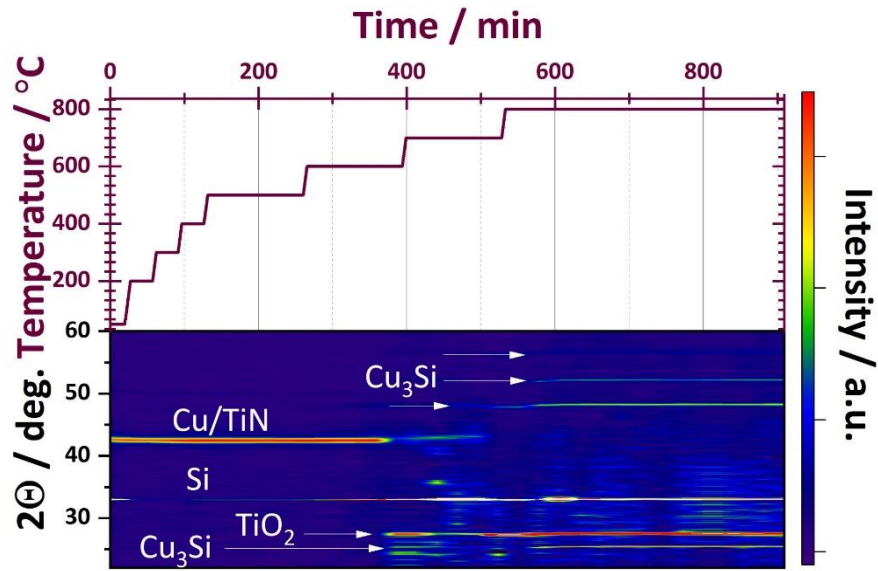

**Fig. S11.** *In situ* temperature- and time-dependent XRD results performed in the atmosphere of $N_2$+5% $H_2$ for a 10 nm-thick TiN thin film grown on Si substrates and covered with a 100 nm-thick Cu thin film.

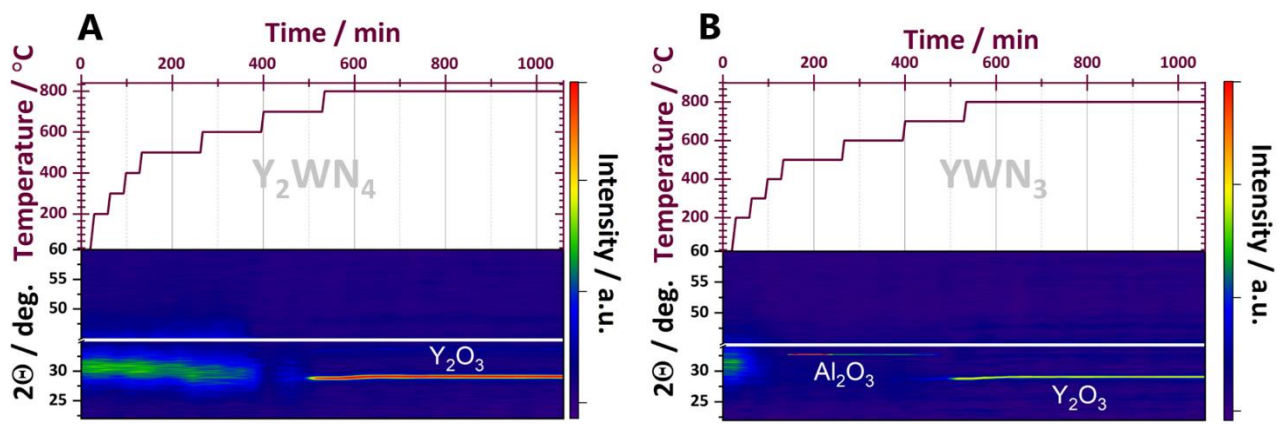

**Fig. S12.** *in situ* temperature- and time-dependent XRD results performed in ambient atmosphere for (A) $Y_2WN_4$ and $YWN_3$ (B) thin film grown on sapphire substrates.